\documentclass[preprint,amsmath,amssymb,nofootinbib]{revtex4}
\usepackage{changes}
\usepackage{amsfonts}
\usepackage{color}
\usepackage[colorlinks=true,
linkcolor=blue,
breaklinks=true,
urlcolor=blue,
citecolor=blue]{hyperref}

\usepackage{amsmath,graphicx,color,epsfig,slashed}

\newcommand{\be}{\begin{equation}}
\newcommand{\ee}{\end{equation}}
\newcommand{\bea}{\begin{eqnarray}}
\newcommand{\eea}{\end{eqnarray}}
\newcommand{\ba}{\begin{array}}
\newcommand{\ea}{\end{array}}
\newcommand{\bi}{\begin{itemize}}
\newcommand{\ei}{\end{itemize}}

\newcommand{\mcb}{{\mathcal B}}

\newcommand{\mcf}{{\mathcal F}}

\renewcommand{\vec}[1]{\mbox{\boldmath $#1 \!\!$ \unboldmath}}

\newcommand{\nslash}{\kern 0.2 em n\kern -0.50em /}
\newcommand{\kslash}{\kern 0.2 em k\kern -0.45em /}
\newcommand{\qslash}{\kern 0.2 em q\kern -0.45em /}
\newcommand{\pslash}{\kern 0.2 em p\kern -0.50em /}
\newcommand{\rslash}{\kern 0.2 em r\kern -0.50em /}
\newcommand{\sslash}{\kern 0.2 em s\kern -0.50em /}
\newcommand{\Sslash}{\kern 0.2 em S\kern -0.50em /}
\newcommand{\Pslash}{\kern 0.2 em P\kern -0.50em /}
\newcommand{\Dslash}{\kern 0.2 em D\kern -0.65em /\kern 0.15em}
\newcommand{\lf}{\left}
\newcommand{\rg}{\right}

\begin{document}
\title{Photoproduction of strange hidden-charm and hidden-bottom states}

\author{Xu Cao{\footnote{caoxu@impcas.ac.cn}}}
\affiliation{Institute of Modern Physics, Chinese Academy of Sciences, Lanzhou 730000, China}
\affiliation{University of Chinese Academy of Sciences, Beijing 100049, China}

\author{Jian-Ping Dai{\footnote{daijianping@ynu.edu.cn}}}
\affiliation{Department of Physics, Yunnan University, Kunming 650091, China}

\author{Zhi Yang{\footnote{zhiyang@uestc.edu.cn}}}
\affiliation{School of Physics, University of Electronic Science and
Technology of China, Chengdu 610054, China}

\begin{abstract}
  Recently BESIII collaboration discovered a charged strange hidden-charm state $Z_{cs}$(3985) in the $D_s^-D^{*0} + D_s^{*-}D^{0}$ spectrum. A higher $Z'_{cs}$ state coupling to $\bar{D}_s^{*-}D^{*0}$ is expected by SU(3)-flavor symmetry, and their bottom partners are anticipated by heavy quark flavor symmetry. Here we study the photoproduction of these exotic states and investigate carefully the background from Pomeron exchange.
  Our results indicate that the maximal photoproduction cross section of strange partner is around 1 $\sim$ 2 orders of magnitude smaller than that of the corresponding non-strange states. The possibility of searching for them in future electron-ion colliders (EIC) is briefly discussed.
\end{abstract}
\maketitle

\section{Introduction} \label{sec:intro}

Exotic hadron candidates challenge our understanding of the strong interaction and hence shed light on the underlying mechanism constrained by Quantum Chromodynamics (QCD)~\cite{Chen:2016qju,Chen:2016spr,Lebed:2016hpi,Esposito:2016noz,Olsen:2017bmm,Guo:2017jvc,Liu:2019zoy,Brambilla:2019esw,Guo:2019twa}.
The states in the charm and bottom region are arranged into the well organized patterns in the spectroscopy, constrained by a high degree heavy quark spin and flavor symmetry (HQSS and HQFS). These patterns apply successfully and evidently to the pentaquark $P_c$ in light of the latest LHCb data~\cite{Aaij:2019vzc}. The scattering potential involving only short-range operators constrained by heavy quark spin symmetry can manifest the first example of full and complete HQSS multiplet of baryon-like $P_c$~\cite{Liu:2019tjn,Xiao:2019aya,Du:2019pij}. The one-pion exchange as well as the long range interaction would also play an important role~\cite{Du:2019pij,Yang:2011wz}. The strange hidden-charm pentaquark $P_{cs}$ was investigated by several theoretical methods~\cite{Wu:2010jy,Wu:2010vk,Feijoo:2015kts,Xiao:2019gjd,Pimikov:2019dyr,Zhang:2020cdi,Peng:2020hql}. Until recently its first evidence was announced by the LHCb Collaboration in the  $J/\psi \Lambda$ invariant mass distribution~\cite{Aaij:2020gdg}.

\begin{figure}
  \begin{center}
  {\includegraphics*[width=0.6\textwidth]{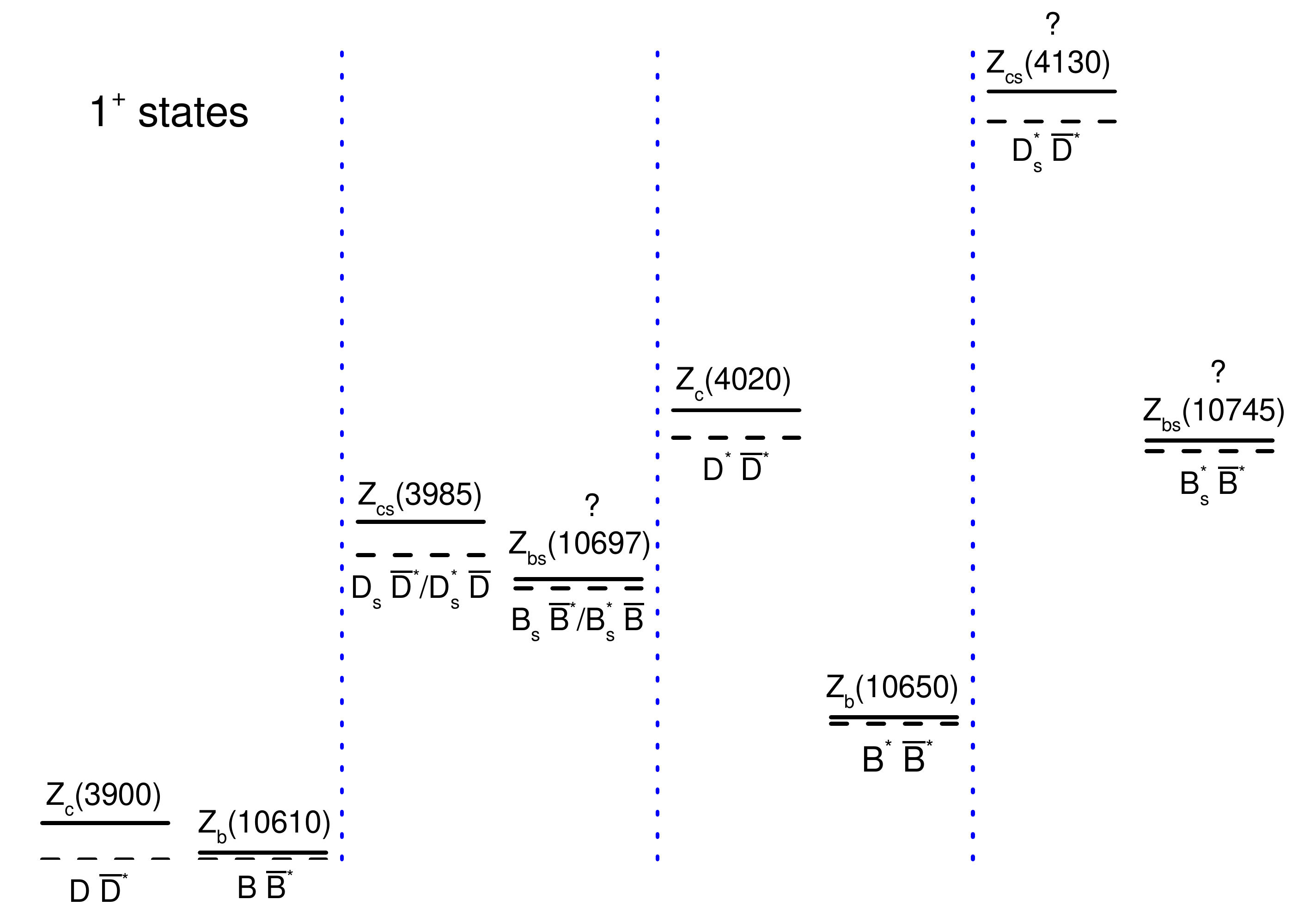}}
    \caption{The axial-vector heavy quark meson $Z$ spectroscopy in the SU(3)-flavor symmetry and HQFS. The $Z_{c/cs}^{(')}$ and $Z_{b/bs}^{(')}$ are non-strange/strange hidden-charm and hidden-bottom states, respectively. The masses in the brackets of the states labeled in interrogation mark are from theoretical expectations~\cite{Yang:2020nrt}, while the others are experimental ones. } 
    \label{fig:Zspectrum}
  \end{center}
\end{figure}

In the meson-like sector, the $Z_{c}$(3900) 
with a width of 28.3$\pm$2.5~MeV\cite{Zyla:2020zbs}, and $Z_{c}$(4020)
with a width of 13$\pm$5~MeV~\cite{Zyla:2020zbs}, locating just above $D \bar{D}^{*}$ and $D^* \bar{D}^{*}$ threshold respectively, are widely considered as the most evident twin molecules with hidden charm under HQSS~\cite{Nieves:2012tt,HidalgoDuque:2012pq,Guo:2013sya,Guo:2017jvc,Tanabashi:2018oca}.
The $Z_{b}$(10610) and $Z_{b}$(10650) are their close analogs in bottom sector under HQFS~\cite{Guo:2013sya}.
Very recently BESIII collaboration~\cite{1830518} released the evidence of a strange hidden-charm state, nominated as $Z_{cs}(3985)^{\pm}$ with a mass of $3982.5^{+1.8}_{-2.6} \pm {2.1} \,{\rm MeV}$ and a narrow width of $12.8^{+5.3}_{-4.4} \pm {3.0} \,{\rm MeV}$ in the $D_s^-D^{*0} + D_s^{*-}D^{0}$ spectrum of $e^+ e^- \to K^+ (D_s^-D^{*0} + D_s^{*-}D^{0})$ at $\sqrt{s} = 4.681$~GeV. It locates around 9.5 $\pm$ 3.7 MeV above $D_s^-D^{*0} + D_s^{*-}D^{0}$ threshold~\cite{Zyla:2020zbs}, close to that for $Z_{c}$(3900) and $Z_{c}$(4020), reflecting the strong constraint from SU(3) flavour symmetry~\cite{Yang:2020nrt}.
This would be the first exotic candidate in strange hidden-charm sector, followed immediately by several theoretical studies~\cite{Yang:2020nrt,Meng:2020ihj,Wang:2020kej,Wan:2020oxt,Wang:2020iqt,Ikeno:2020csu}. Especially, its HQSS partner, $Z_{cs}^{\prime}$, is predicted in Ref.~\cite{Yang:2020nrt,Meng:2020ihj}. Together with the very recent $X_0$(2900)/$X_1$(2900) with quark content $\bar{c} \bar{s} u d$ by LHCb~\cite{Aaij:2020hon,Aaij:2020ypa}, the discovery of $Z_{cs}(3985)$ state extends the study of exotic hadrons to strange sector. They would constitute the first full HQSS multiplet of charmonium-like states together with the non-strange hidden-charm states~\cite{Guo:2009id,Xiao:2019aya,Xiao:2019gjd,Xiao:2013yca,Mehen:2011yh,Valderrama:2012jv,Nieves:2012tt,HidalgoDuque:2012pq,Guo:2013sya,Guo:2013xga,Baru:2016iwj,Liu:2019stu,Liu:2018zzu}, as shown in Fig.~\ref{fig:Zspectrum}. The locations of the states in bottom sector are closer to the corresponding thresholds than those in charm sector, since the binding energy is roughly inversely proportional to quark mass. 
As the strange partner of $Z_c$(3900), the assignment of its quantum numbers is expected to be $I(J^P) = 1/2~(1^+)$.

A particular feature in the strange sector is that the one-pion exchange is forbidden by parity and isospin in the $D_s^-D^{*0}/D_s^{*-}D^{0}$ interaction. Other mesons, e.g. $\eta^{(')}$, $\sigma$, $\phi$ and $\omega$, are allowed but the effective interaction range is less than 0.5 fm, much shorter than $\pi$-exchange~\cite{Sun:2011uh,Valderrama:2019chc}. Whether the induced strength by these heavier meson is enough to bind the $D_s^-D^{*0}/D_s^{*-}D^{0}$ system is questioned because of the OZI suppression~\cite{Aceti:2014uea,Dias:2014pva}. Considering that the bound between two strange heavy mesons is much tighter than that for non-strange case, the underlying scenario would be dominant by short range contact interaction as discussed in Ref.~\cite{Peng:2020xrf}. 
Thus the mechanism for its formation needs further investigation.
The $J/\psi K$ decay channel of $Z_{cs}$ was thought to be important in the earlier studies, e.g. QCD sum rule~\cite{Lee:2008uy,Dias:2013qga}, initial $K$-meson emission mechanism~\cite{Chen:2013wca} and the hadrocharmonium picture~\cite{Voloshin:2019ilw}. A rich spectrum of $Z_{cs}$ and $Z_{bs}$ coupled to the hidden channels was also predicted in the compact tetraquark~\cite{Ferretti:2020ewe}. However, null results in $e^+ e^- \to K^+ K^- J/\psi$ channel from Belle~\cite{Shen:2014gdm,Yuan:2007bt}
and BESIII~\cite{Ablikim:2018epj} indicate a relatively small $J/\psi K$ partial decay width, calling for higher statistics in future BelleII and BESIII experiments.

The photoproduction of some exotic states, e.g. $Z_c$(4430) and pentaquarks $P_{c,b}$, have attracted a lot of theoretical attention~\cite{Liu:2008qx,Galata:2011bi,Wang:2015jsa,Karliner:2015voa,Kubarovsky:2015aaa,Huang:2016tcr,Blin:2016dlf,Cao:2019kst,Wu:2019adv,Wang:2019krd,Cao:2019gqo,Xie:2020niw,Yang:2020eye}, due to the possible impact of kinematic effect on the resonance structure~\cite{Rosner:2006vc,Guo:2015umn}. Particularly, though the triangle diagrams could be present in photoproduction, it is hardly possible to satisfy the on-shell condition of the triangle singularity~\cite{Liu:2016dli}, thus they are expected to be negligible. If the experiments are motivated to find their photoproduction, the kinematic explanation, e.g. kinematic reflection or triangular sigularity, will be definitely excluded and these structures are surely real states. On the other hand, the discovery potential in photoproduction is also very essential. For example, the exotic $Y$(3940)~\cite{He:2009yda}, $X$(3915)~\cite{Lin:2013ppa}, $Z_c$(3900)~\cite{Lin:2013mka}, and $Z_c$(4200)~\cite{Wang:2015lwa} through photoproduction were discussed theoretically. Experimental efforts were also devoted to the photoproduction of $P_c$~\cite{Ali:2019lzf} by GlueX collaboration,  $Z_c$(3900)~\cite{Adolph:2014hba} and $X$(3872)~\cite{Aghasyan:2017utv} by COMPASS collaboration. 
In this paper, we study the $Z_{cs/bs}^{(')}$ photoproduction through $\gamma p \to \Lambda Z_{cs/bs}^{(')}$ processes, with their subsequent decaying into hidden-charm/hidden-bottom channels.

%

\section{Formalism and results} \label{sec:Frwk}
\begin{figure}
  \begin{center}
  {\includegraphics*[width=0.6\textwidth]{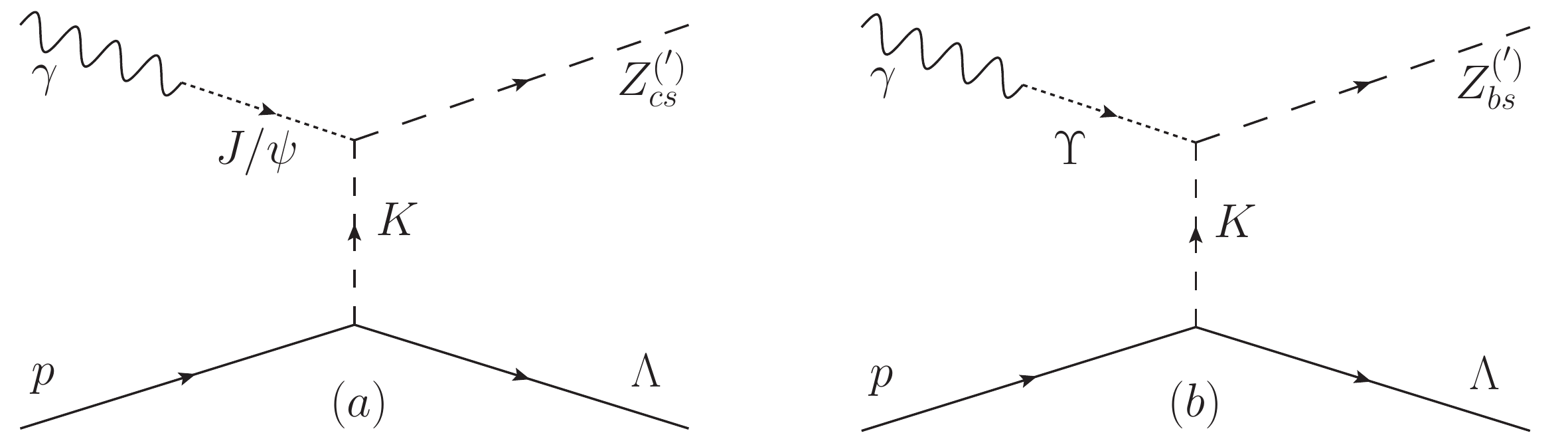}}
    \caption{The $Z_{cs/bs}^{(')}$ photoproduction in the process of (a) $\gamma p \to \Lambda Z_{cs}^{(')}$ and (b) $\gamma p \to \Lambda Z_{bs}^{(')}$, respectively. }
    \label{fig:Zbcsphoto_fey}
  \end{center}
\end{figure}
\begin{figure}
  \begin{center}
  {\includegraphics*[width=0.9\textwidth]{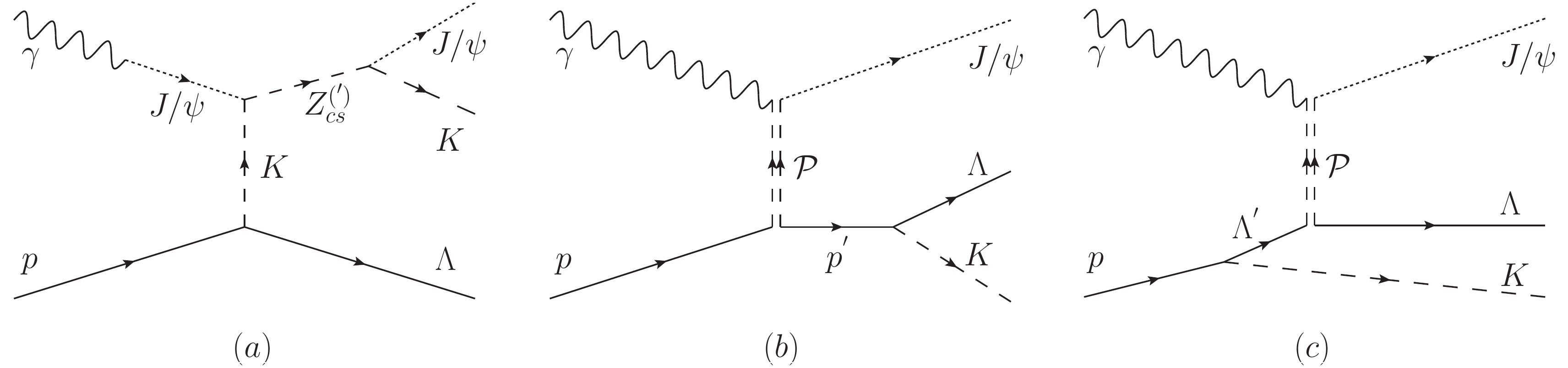}}
    \caption{The diagrams of $\gamma p \to V K \Lambda$ reaction. (a): The $Z_{cs}^{(')}$ photoproduction in $\gamma p \to J/\psi K \Lambda$ reaction with $Z_{cs}^{(')}$ decaying to $J/\psi K$ channels. (b) and (c): Background contribution through Pomeron exchange. Similar diagrams for $\gamma p \to \Upsilon K \Lambda$ can be obtained by substituting $\Upsilon$ for $J/\psi$ and $Z_{bs}^{(')}$ for $Z_{cs}^{(')}$.}
    \label{fig:Zbcsphoto3}
  \end{center}
\end{figure}
\begin{table}
  \begin{center}
 \begin{tabular}{c|c|c|c|c|c}
\hline\hline
  $Z$ states   & Mass $M_Z$ (MeV) & Width $\Gamma_Z$ (MeV) & $VM$ channel   & $\mcb(Z \to VM)$  & $g_{ZVM}$   \\
\hline
  $Z_{cs}$     &     3985.0       &    13.0         & $J/\psi K$            & 1.0$^\dag$           & 3.2$\sqrt{\mcb}$   \\
\hline
  $Z_{cs}^{'}$ &     4130.0       &    13.0         & $J/\psi K$            & 1.0$^\dag$            & 2.6$\sqrt{\mcb}$    \\
\hline
  $Z_{bs}$     &                  &                 & $\Upsilon(1S) K$      & 1.0$^\dag$            & 3.8$\sqrt{\mcb}$    \\
               &     10697.0      &    10.0         & $\Upsilon(2S) K$      & 1.0$^\dag$            & 10.4$\sqrt{\mcb}$    \\
\hline
  $Z_{bs}^{'}$ &                  &                 & $\Upsilon(1S) K$      & 1.0$^\dag$            & 3.6$\sqrt{\mcb}$    \\
               &     10745.0      &    10.0         & $\Upsilon(2S) K$      & 1.0$^\dag$            & 9.1$\sqrt{\mcb}$    \\
\hline
\hline
  $Z_{b}$      &                  &                 & $\Upsilon(1S)\pi$     & 0.54\%         & 0.49   \\
               &     10607.2      &    18.4         & $\Upsilon(2S)\pi$     & 3.6\%          & 3.3    \\
               &                  &                 & $\Upsilon(3S)\pi$     & 2.1\%          & 9.3    \\
\hline
  $Z_{b}^{'}$  &                  &                 & $\Upsilon(1S)\pi$     & 0.17\%         & 0.21   \\
               &     10652.2      &    11.5         & $\Upsilon(2S)\pi$     & 1.4\%          & 1.5    \\
               &                  &                 & $\Upsilon(3S)\pi$     & 1.6\%          & 4.9    \\
\hline\hline
   \end{tabular}
  \end{center}
  \caption{The parameters of $Z_{cs}^{(')}$, $Z_{bs}^{(')}$ and $Z_{b}^{(')}$ used in our calculation.  Note that $Z_{bs}^{(')}$ are below $\Upsilon(3S) K$ mass threshold. The parameters of $Z_{b}^{(')}$ and $\mcb(Z_{b}^{(')} \to V \pi) $ are adopted from PDG~\cite{Zyla:2020zbs}. \\
  $^\dag$: Because there are no measurements on the branching fractions of $Z^{(\prime)}_{cs}$ and $Z^{(\prime)}_{bs}$ decaying to $VK$, their couplings $g_{ZVK}$ are computed with the full width of the corresponding states, e.g. $\mcb(Z \to VK) = $ 1.0 with $V$ being the vector heavy quarkonium.
  \label{tab:Zcouplings}}
\end{table}

The $\gamma p \to \Lambda Z_{cs/bs}^{(')}$ reactions can be proceeded by one-kaon exchange as shown in Fig.~\ref{fig:Zbcsphoto_fey}, where $Z_{cs/bs}^{(')}$ then decay into vector quarkonium and kaon, e.g. $J/\psi K$ or $\Upsilon K$~\cite{Dias:2013qga,Chen:2013wca,Ferretti:2020ewe}.

In the vector meson dominant (VMD) model~\cite{Bauer:1977iq,Meissner:1987ge}, the coupling of a vector meson $V$ with mass $M_V$ to photon $\gamma$ is written as
\bea
{\cal L}_{V\gamma} = - \frac{eM_V^2}{f_V}V_\mu A^\mu \, ,
\eea
where $f_V$ is the decay constant of the vector meson, determined by the dilepton decay width $\Gamma_{V\to e^{+}e^{-}}$,
\bea \label{eq:vmd}
\frac{e}{f_V} &=& \left[\frac{3\Gamma_{V\to e^+ e^-}}{\alpha M_V}\right]^{1/2}
\eea
with $\alpha = e^2/(4\pi) = 1/137$ being the fine-structure constant.
We obtain $f_{J/\psi}= 11.16$, $f_{\Upsilon(1S)}= 39.69$, $f_{\Upsilon(2S)}= 60.46$, and $f_{\Upsilon(3S)}= 72.23$ with the $M_V$ and $\Gamma_{V\to e^+ e^-}$ from the Particle Data Group (PDG)~\cite{Zyla:2020zbs}.
While the effective Lagrangian for the coupling of $N K \Lambda$ is taken from Ref.~\cite{Cao:2010km} as
\bea
{\cal L}_{N K \Lambda} &=& -i g_{N K \Lambda} \bar{N}\gamma_5 K \Lambda + h.c. \, ,
\eea
with the coupling constant $g_{N K \Lambda} = 14.0 $~\cite{Goldberger:1958tr}.

As discussed in Sec.~\ref{sec:intro}, the S-wave $D_s^-D^{*0} + D_s^{*-}D^{0}$ molecular state~\cite{Yang:2020nrt,Meng:2020ihj,Wan:2020oxt,Ikeno:2020csu} and compact tetraquark~\cite{Wan:2020oxt,Wang:2020iqt} are the most popular explanations for $Z_{cs}$, both of which expect the spin-parity $J^P=1^{+}$.
In the following formula, we use $Z$ and $V$ to denote $Z_{cs/bs}^{(')}$ and vector quarkonium, respectively. Then the effective Lagrangian of the $Z V K$ coupling is~\cite{Liu:2008qx}
\begin{eqnarray}
\mathcal {L}_{Z V K} &=& \frac{g_{ZVK}}{M_Z} \, \partial^\mu V^\nu \, \lf( \partial_{\mu}\vec{P} \cdot \vec{Z}_{\nu} - \partial_{\nu}\vec{P} \cdot \vec{Z}_{\mu} \rg).
\end{eqnarray}
Note that above effective Lagrangian depends on the quantum number rather than the internal components of $Z_{cs}$. The information on the internal structure is coded in the coupling $g_{ZVK}$ and can be inputted from different models of $Z_{cs/bs}^{(')}$. On the other hand, the measured cross sections can be used to determine $g_{ZVM}$ by comparing with our results. Eventually this is helpful for giving the information on the internal structure. For this purpose, the dimensionless coupling constant $g_{ZVK}$ can be related to the corresponding decay width
\bea \label{eq:gamma-z}
\Gamma\left(Z \to VK\right) &=&\left(\frac{g_{ZVK}}{M_Z}\right)^2 \frac{ |{\vec{p}_{cm}}|}{16\pi M_Z^2}
 \left[2 (M_Z E_{cm} - M_K^2)^2 + M_{V}^2 E_{cm}^2 \right] ,
\eea
where $\vec{p}_{cm}$ and $E_{cm}$ are the three-vector momentum and energy of the $K$-meson in the $Z$ rest frame, respectively. The masses, widths and the coupling strengths of $Z^{(\prime)}$ are summarized in Table~\ref{tab:Zcouplings}, where the branching ratios of $Z^{(\prime)}_{cs,bs}$ are assumed to be 1.0. So the final calculated cross sections are in fact $\sigma(\gamma p \to \Lambda Z)/\mcb(Z \to VK)$ and $\sigma(\gamma p \to \Lambda V K)/\mcb(Z \to VK)$. We introduce a form factor $\mcf_{s}$ to cut off the contributions of large four-momenta from the off-shell $Z$ resonance,
\be \label{eq:ffZ}\mcf_{s}(\Lambda_Z,M_Z) = \frac{\Lambda_Z^4}{\Lambda_Z^4 + (q_Z^2-M_Z^2)^2} \, .
\ee
Here the cutoff is set to be the mass of the intermediate vector meson, e.g. $\Lambda_Z = M_{\psi(nS)} $ or $M_{\Upsilon(nS)}$~\cite{Friman:1995qm}. An alternative form factor is usually used in literature,
\be
\mcf_{Z}(\Lambda_Z,M_Z) = \frac{\Lambda_Z^2 - M_Z^2}{\Lambda_Z^2 - q_Z^2}  \, ,
\ee
which will not change the near-threshold results in charm sector much because we concentrate on the energies where $Z_{cs}^{(')}$ is nearly on-shell. While for $Z_{bs}^{(')} \to \Upsilon(2S) K$ vertex, this form factor encounters a non-physical pole~\cite{Cao:2010km}, which shall be avoided. Thus the form factor in Eq.~(\ref{eq:ffZ}) is a more reasonable choice. A monopole form factor is introduced to suppress the contributions from high exchanged momenta of the off-shell $t$-channel propagator,
\be \label{eq:tff}
\mcf_{t}(\Lambda_t,m_t) = \frac{\Lambda_t^2 - m_t^2}{\Lambda_t^2 - t} \, ,
\ee
where $\Lambda_t$ is the corresponding cutoff parameter.

With the above prescription, we obtain the differential cross section of $\gamma p \to Z_{cs}/Z_{bs} \Lambda$ in Fig.~\ref{fig:Zbcsphoto_fey} as
\be
\frac{\textrm{d} \sigma}{\textrm{d} t} = \frac{1}{64\pi s} \frac{1}{|\vec{k}_{cm}|^2}
 \frac{1}{4} \left(g_{NK\Lambda} \frac{g_{ZVK }}{M_Z}
\frac{e}{f_{V}}\right)^2 \frac{((m_\Lambda-M_N)^2-t) (M_Z^2-t)^2}{(m_K^2-t)^2}
\mcf_{t}^2(\Lambda_K,m_K) \mcf_{t}^2(\Lambda_Z,m_K),
\ee
where $t$ is the square of the four-momentum transfer from initial proton to final $\Lambda$-baryon, $s$ is the c.m. energy square and $|\vec{k}_{cm}| = (s-m_N^2)/2\sqrt{s}$ is the photon energy in the c.m. frame. For the cut-off values, we use $\Lambda_K =$ 0.7 GeV and $\Lambda_Z = M_V$. At high energies an alternative Regge propagator from $K$-meson Regge trajectory would be substituted for the usual meson propagator~\cite{Guidal:1997hy}. This has been adopted in the photoproduction of $Z_c$(4430)~\cite{Galata:2011bi} and $Z_c$(4200)~\cite{Wang:2015lwa}. Since we concentrate on the threshold region, the choice of ordinary meson propagator in above equation is favored.

The amplitude of $\gamma p \to V K \Lambda$ in Fig.~\ref{fig:Zbcsphoto3} reads generally as
\bea \label{eq:amp3}
\mathcal{T}_{fi} &=& \epsilon_{\gamma\mu}\epsilon_{V \nu}^\ast \lf( \mathcal{M}_Z^{\mu \nu} + \mathcal{M}_\mathcal{P}^{\mu \nu} \rg),
\eea
where $\epsilon_{\gamma\mu}$ and $\epsilon_{V \nu}$ are the polarization vector of initial photon $\gamma$ and final vector quarkonium $V$, respectively. Hereafter we take $\gamma p \to J/\psi K \Lambda$ as an example. The amplitude of $\gamma p \to \Lambda \Upsilon K$ can be directly obtained with the replacement of intermediate strange states and final vector quarkonium by those in bottom sector. The transition tensor of $\gamma p \to Z_{cs}^{(')} \Lambda \to J/\psi K \Lambda$ in Fig.~\ref{fig:Zbcsphoto3}(a) is
\bea
\mathcal{M}_Z^{\mu \nu} &=& -i \, g_{N K \Lambda} \left(\frac{g_{ZVK }}{M_Z}\right)^2 \frac{e}{f_{V}} \, \bar{u}_{\Lambda}(p{'})\gamma_5 u_p(p) \nonumber\\ && \times\left((p{'} - p)\cdot q \, g^{\mu\beta}-(p{'} - p)^\beta q^\mu\right)\left(p_V \cdot p_K g^{\alpha\nu}- p_V^\alpha p_K^\nu\right) \nonumber \\
&& \times \frac{1}{t-m_K^2} \frac{-g_{\alpha\beta}+p_{Z\alpha}p_{Z\beta}/M_Z^2}{p_Z^2-M_Z^2+i M_Z\Gamma_Z} \mcf_{t}(\Lambda_K,m_K) \mcf_{t}(\Lambda_Z,m_K) \mcf_{s}(\Lambda_Z,M_Z),
\eea
where $q$, $p$, $p{'}$, $p_V$, $p_K$ and $p_Z = p_V + p_K$ are the momentum of photon, proton, $\Lambda$, $J/\psi$, $K$-meson and $Z_{cs}^{(')}$, respectively. Because $Z_{cs}^{(')}$ and $Z_{bs}^{(')}$ states are narrow, the cross sections in Fig.~\ref{fig:Zbcsphoto3}(a) can be approximately estimated by $\sigma(\gamma p \to Z_{cs}^{(')} \Lambda) \mcb(Z_{cs} \to J/\psi K)$.

The Pomeron contribution is the main background in the search of $Z_{cs}^{(')}/Z_{bs}^{(')}$ signal. This diffractive process with a Pomeron trajectory $\mathcal{G}_P(s,t')= -i(\alpha^\prime s)^{\epsilon+\alpha^\prime t'}$ could be proceeded by final and initial emission of $K$-meson, as depicted respectively in Fig.~\ref{fig:Zbcsphoto3}(b) and Fig.~\ref{fig:Zbcsphoto3}(c). The parameters of Pomeron trajectory are well known, e.g. $\epsilon$ = 0.08 and $\alpha^\prime$ = 0.25 GeV$^{-2}$~\cite{Laget:1994ba,Guidal:1997hy}. The $\gamma V \mathcal{P}$ vertex can be described by gauge invariant coupling $2\beta_c V(t{'}) T_{\mu\alpha\nu}$~\cite{Titov:1998bw,Zhao:1999af} with
\bea
T^{\mu\alpha\nu} &=& (q + k_V)^\alpha g^{\mu\nu} -2q^\nu g^{\alpha\mu}, \\
V(t') &=& \frac{ 4\mu_0^2}{(M_{V}^2-t{'})(2\mu_0^2+M_{V}^2 -t{'})}  \, ,
\eea
here $\mu_0 = 1.2$ GeV and $\beta_c^2 = 0.8$ GeV$^2$ for charmonium, and $\beta_b^2 = 0.1$ GeV$^2$ for bottomonium.
The Pomeron-nucleon interaction in Fig.~\ref{fig:Zbcsphoto3}(b) can be written in a manner of vector coupling $3\beta_0 f(t{'})\gamma_\mu$, where $\beta_0 = 2$ GeV is the coupling constant between Pomeron and the constituent quark within nucleon. The $f(t{'})$ is the parameterized nucleon electromagnetic form factor (EFF)~\cite{Laget:1994ba}
\be
f(t{'}) = \frac{4M_N^2-2.8t{'}}{(4M_N^2-t{'})(1-t{'}/0.7)^2}
\ee
with the squared energy transfer $t{'} = (p_V-q)^2$ in unit of GeV$^2$. This prescription is widely used in $J/\psi$ photoproduction and describes well the data in a wide range of energies~\cite{Wu:2012wta,Wu:2013xma}. Then the amplitude in Fig.~\ref{fig:Zbcsphoto3}(b) is calculated as
\bea
\mathcal{M}_\mathcal{P}^{\mu \nu} &=& 6 \beta_0 \beta_c g_{N K \Lambda}\, f(t')\, V(t')\, \mathcal{G}_P(s,t')\, T^{\mu\alpha\nu}\,  \bar{u}_\Lambda(p_)\, \gamma_{5}\, \frac{\slashed{p}_N' + M_N}{p_N'^2 - M_N^2} \,\gamma_{\alpha} \, u_p(p) \,\mcf_{t'}(\Lambda_N,M_N),
\eea
where $p_N'$ and $M_N$ are the four momentum and mass of intermediate proton, respectively. The form factor $\mcf_{t'}(\Lambda_N,M_N)$ in the form of Eq.~(\ref{eq:tff}) takes into account the off-shell effect of nucleon with $\Lambda_N$ being the cut-off parameter. Other nucleon excited states $N^*$ with strong coupling to $K \Lambda$ would also contribute to this diagram. We neglect them for the moment because of the unknown coupling strength of Pomeron-$N^*$ interaction.
The Pomeron-$\Lambda$ interaction is expected to be similar to that of Pomeron-nucleon. However, both its coupling strength $\beta_0$ and the $\Lambda$ EFF~\cite{Cao:2018kos} are poorly known. Hence here we neglect the contribution in Fig.~\ref{fig:Zbcsphoto3}(c). As a matter of fact, the initial emission of $K$-meson is kinematically unfavored and is expected to be small, similar to the initial emission of $\pi$-meson in the study of various $Z_c$ photoproduction~\cite{Liu:2008qx,Lin:2013mka,Wang:2015lwa}.

We can evaluate the cross sections and distributions of final particles in three-body phase space using the amplitudes in Eq.~(\ref{eq:amp3}). In order to compare with the results in the bottom sector, we also present the results of $\gamma p \to n Z_b^{(')}$, which has been estimated by JPAC~\cite{Albaladejo:2020tzt}. The formalism is quite analogous to above one with the exception of isospin factors, which can be easily found in references~\cite{Liu:2008qx,Lin:2013mka}. The relevant parameters of $Z_{b}^{(')}$ are adopted from PDG~\cite{Zyla:2020zbs} and listed in Table~\ref{tab:Zcouplings}.

\section{Numerical results and discussion} \label{sec:Nmrk}

\begin{figure}
  \begin{center}
  {\includegraphics*[width=0.45\textwidth]{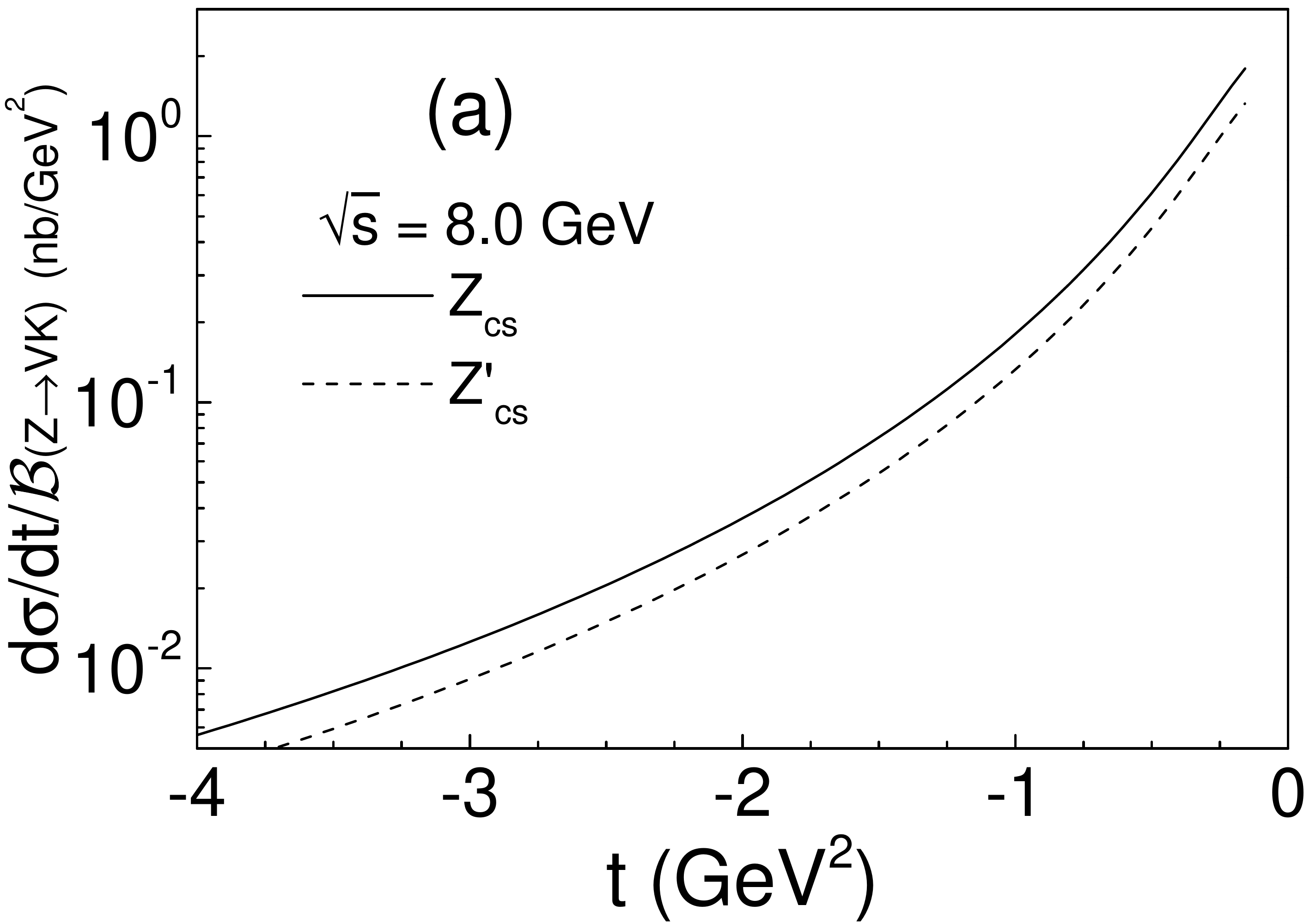}
   \includegraphics*[width=0.45\textwidth]{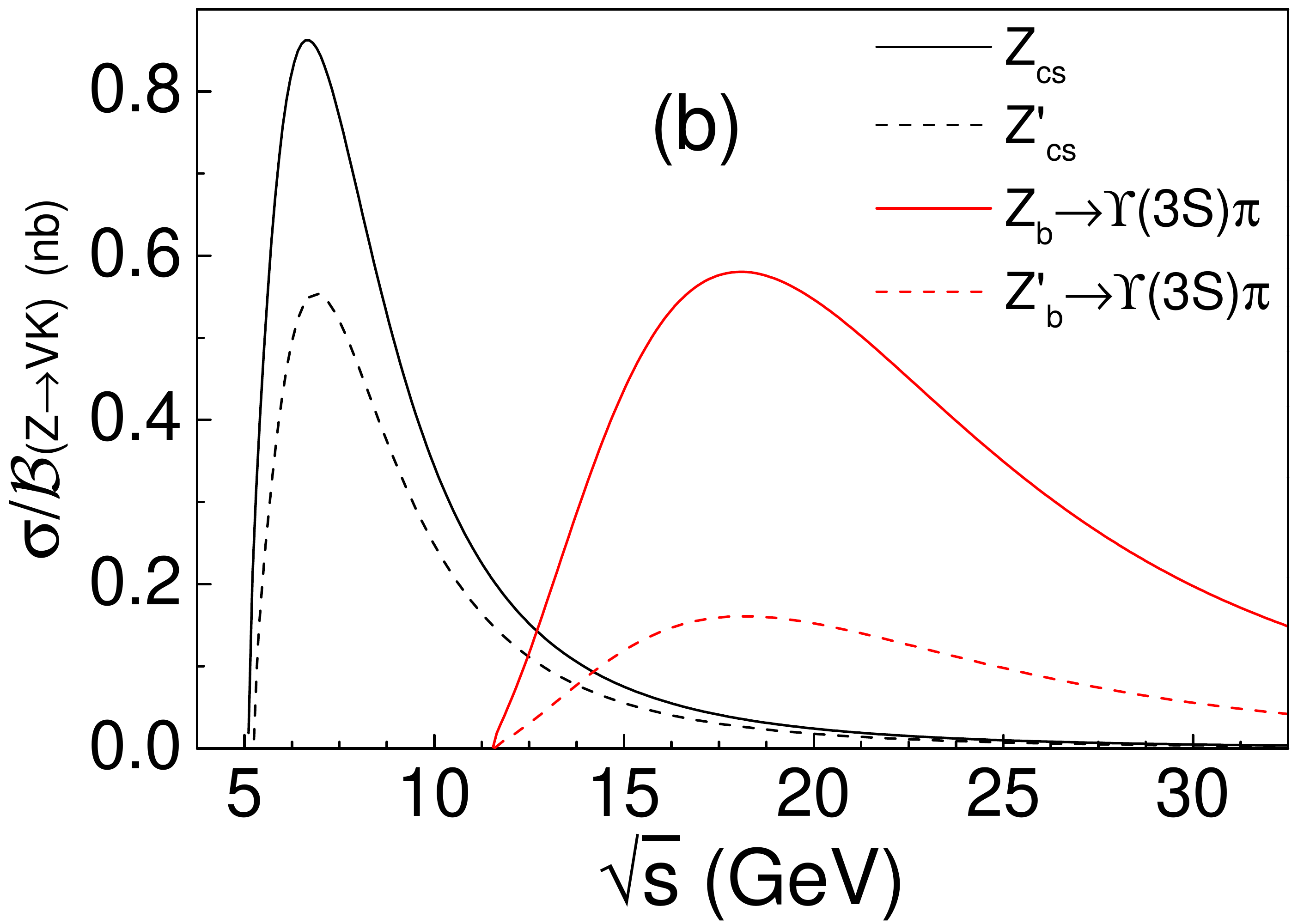}}
    \caption{The differential (a) and total (b) cross sections of the $\gamma p \to \Lambda Z_{cs}^{(')}$. The black solid and dashed lines correspond to results of $Z_{cs}$ and $Z_{cs}^{'}$, respectively. In plot (b) the $\gamma p \to \Lambda Z_{bs}^{(')}$ are also shown for comparison with considering $\Upsilon(3S)$ only in the VMD model. It shall be noted that $\mcb(Z_{cs}^{'} \to J/\psi K)$ are set to be 1.0 while realistic $\mcb(Z_b^{'} \to \Upsilon \pi)$ have been included as shown in Table.~\ref{tab:Zcouplings}.}
    \label{fig:Zcsphoto}
  \end{center}
\end{figure}
\begin{figure}
  \begin{center}
  {\includegraphics*[width=0.45\textwidth]{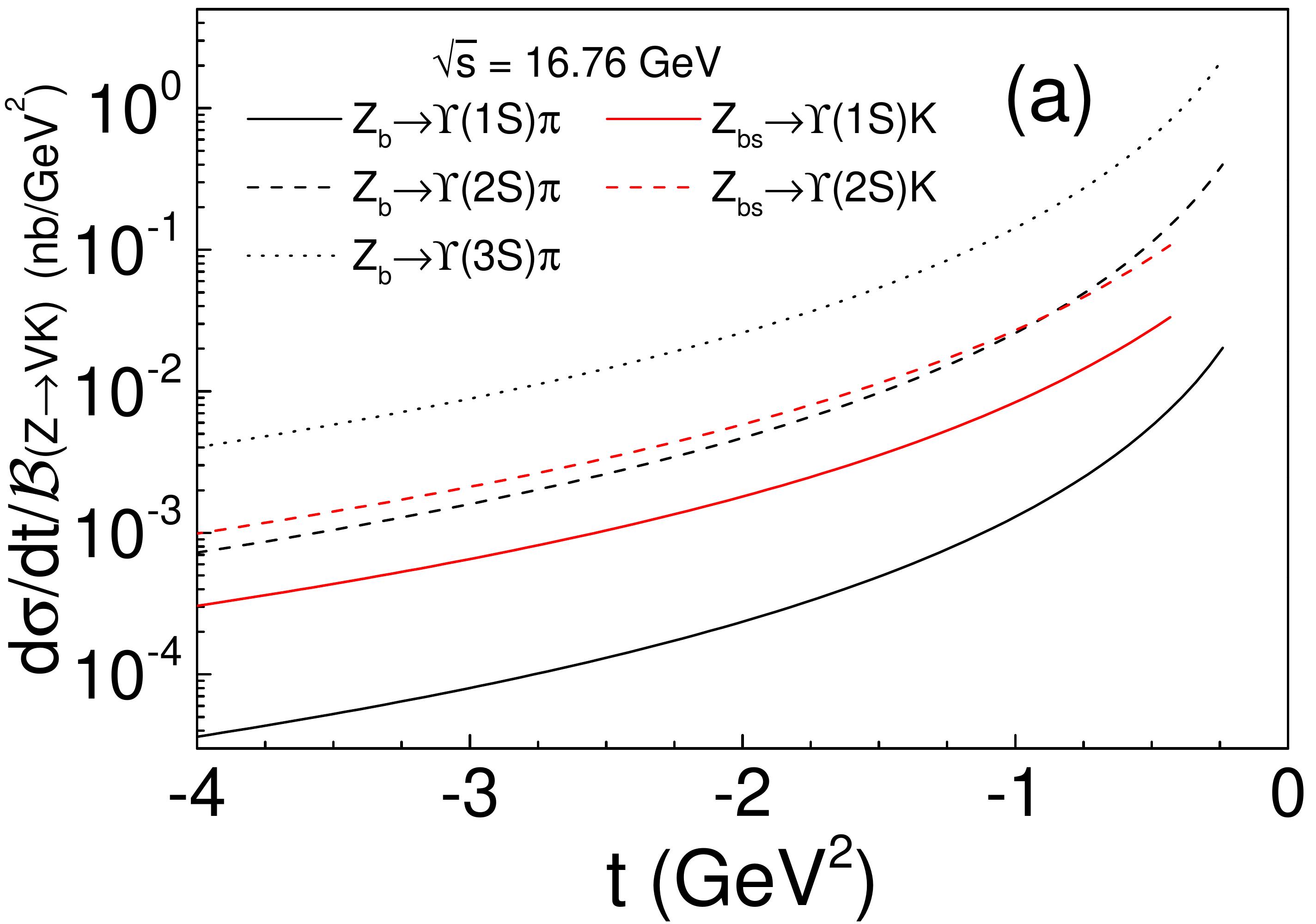}
   \includegraphics*[width=0.45\textwidth]{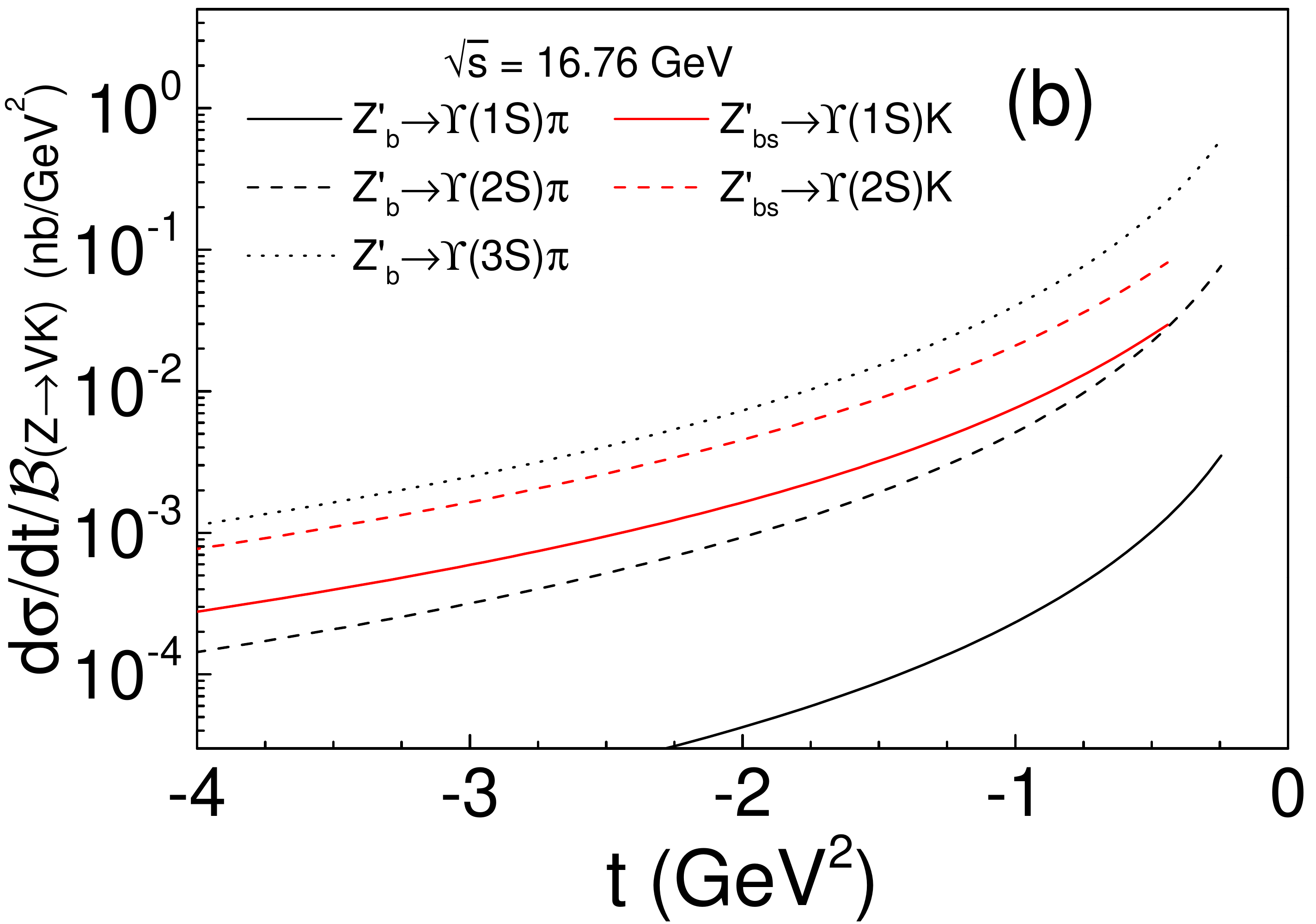}
   \includegraphics*[width=0.45\textwidth]{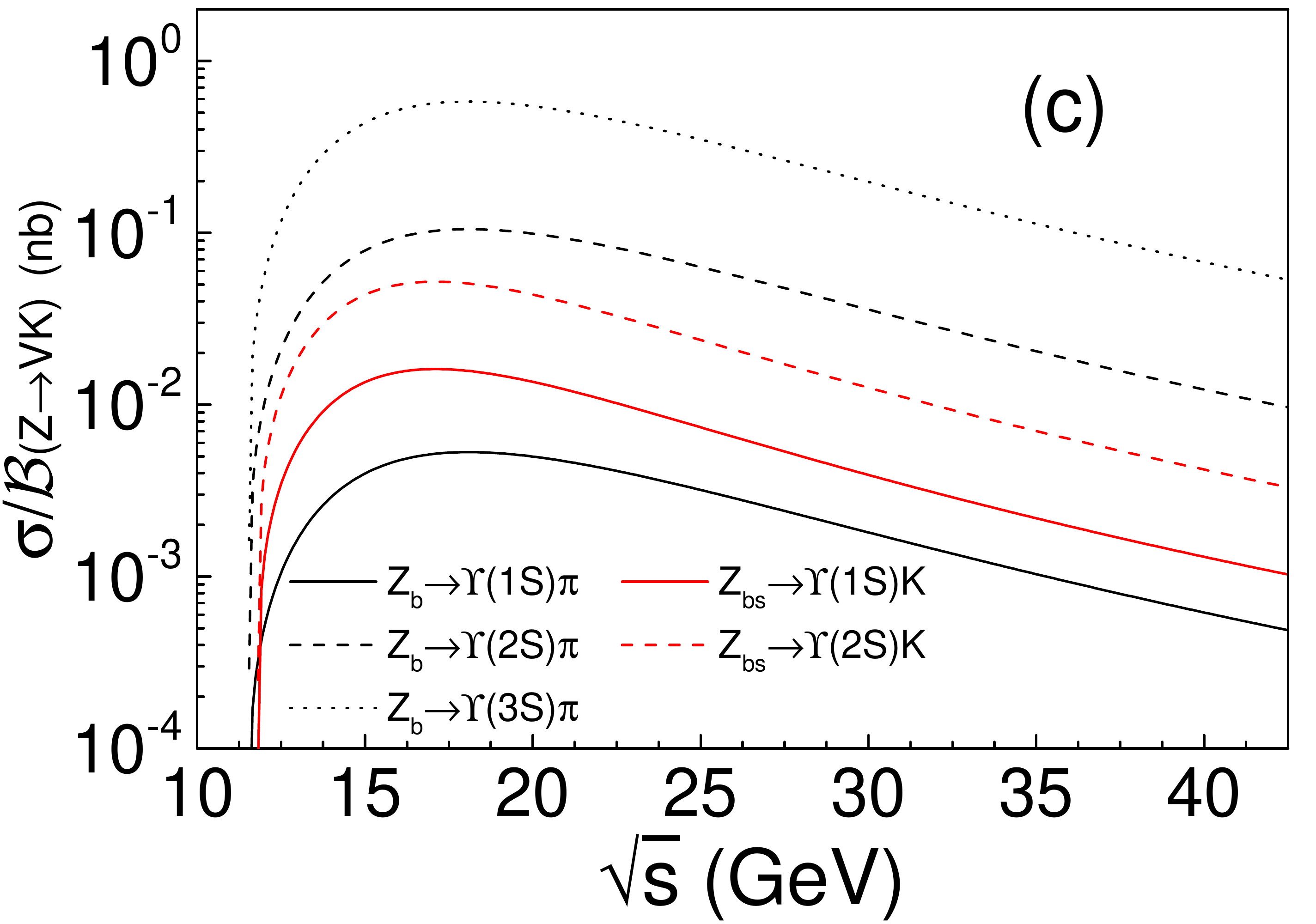}
   \includegraphics*[width=0.45\textwidth]{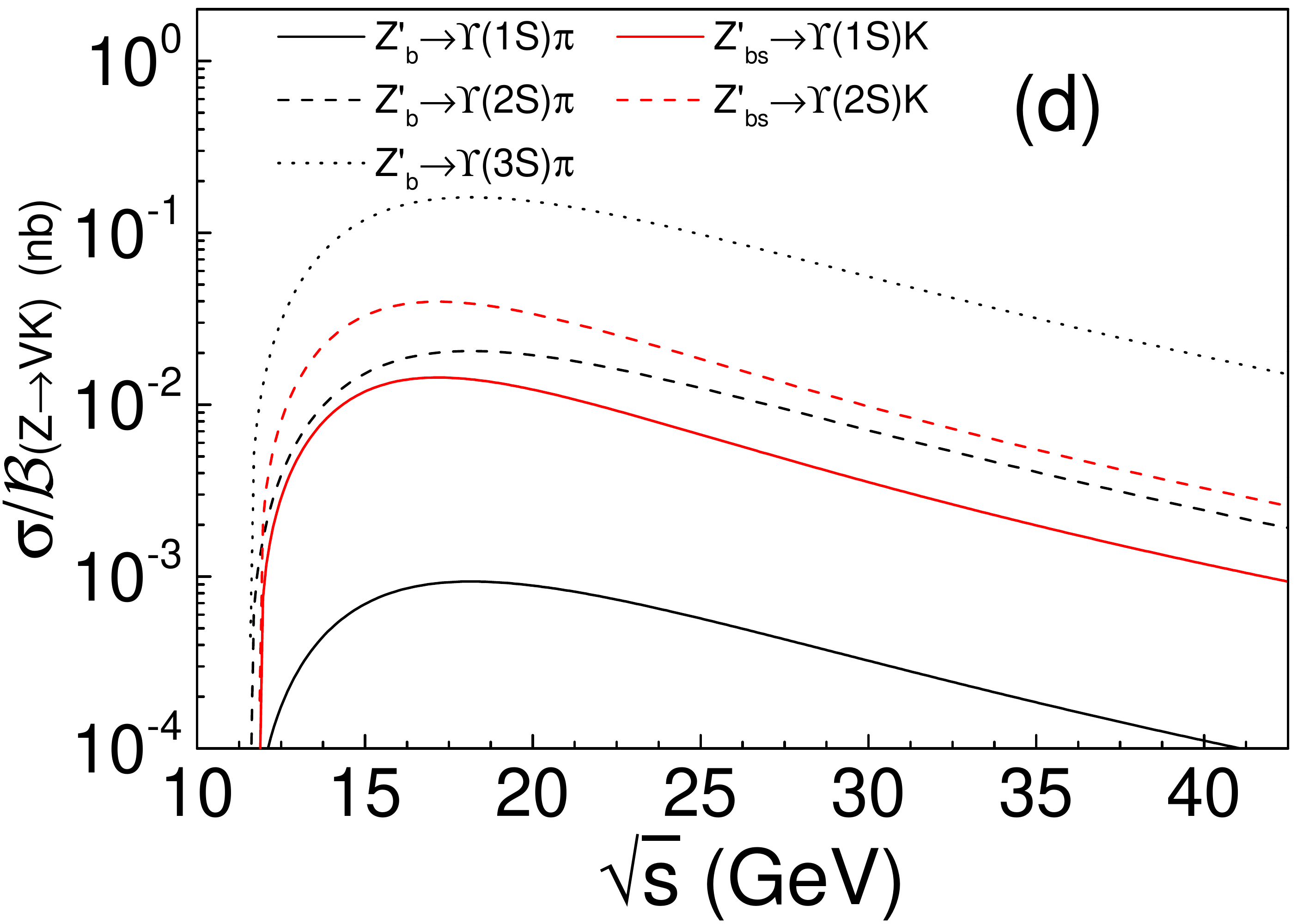}}
    \caption{The differential (top) and total (bottom) cross sections of $\gamma p \to n Z^{(')}_{b}$ and $\gamma p \to \Lambda Z_{bs}^{(')}$, respectively. The black solid, dashed and dotted lines correspond to results of $\gamma p \to n Z_{b}$ with taking into account $\Upsilon(1S)$, $\Upsilon(2S)$, and $\Upsilon(3S)$ in the VMD model, respectively. The red solid and dashed lines are results of $\gamma p \to \Lambda Z_{bs}^{(')}$ with taking into account $\Upsilon(1S)$, and $\Upsilon(2S)$ in the VMD model, respectively. It shall be noted that $\mcb(Z_{bs}^{(')} \to \Upsilon K)$ are set to be 1.0 but realistic $\mcb(Z_b^{(')} \to \Upsilon \pi)$ have been included as shown in Table.~\ref{tab:Zcouplings}.}
    \label{fig:Zbcsphoto}
  \end{center}
\end{figure}

We show the cross sections of $\gamma p \to \Lambda Z_{cs}^{(')}$ in Fig.~\ref{fig:Zcsphoto}.
As explained in Sec.~\ref{sec:Frwk}, we plot the $\sigma / \mcb (Z \to J/\psi K)$ because of the unknown $\mcb (Z \to J/\psi K)$.
The differential cross sections in Fig.~\ref{fig:Zcsphoto}(a) are featured by a typical behavior of $t$-channel meson exchange, decreasing rapidly with larger $|t|$. So the $Z_{cs}^{(')}$ are produced in the forward beam direction. This is also applicable to $Z_{c}^{(')}/Z_{b}^{(')}/Z_{bs}^{(')}$ states, since they are driven by the similar production mechanism. The magnitude of $Z_{cs}^{(')}$ production cross sections are two orders of magnitude smaller than those of the $Z_{c}^{(')}$ states~\cite{Lin:2013ppa,Lin:2013mka}, as can be seen from Fig.~\ref{fig:Zcsphoto}(b). Here we need to point out that for $Z^{(\prime)}_b$ states, since the coupling constants $g_{Z_b\Upsilon \pi}$ are calculated with the measured branching fractions, the cross section distributions in Fig. 4 (b) and Fig. 5 do not divide by ${\cal B}(Z_b \to \Upsilon \pi)$. Because of the close masses, the maximal production cross sections of $Z_{cs}$ and $Z_{cs}^{'}$ are nearly the same and both locate around 7.0 GeV.

We exhibit in Fig.~\ref{fig:Zbcsphoto} the differential and total cross sections of $\gamma p \to B Z^{(')}_{b/bs}$, with $B=n\; \text{and}\; \Lambda$. The contributions of $\Upsilon(1S)$, $\Upsilon(2S)$ and $\Upsilon(3S)$ in the VMD model are separately indicated due to the unknown interference phase between them. It is seen that $\Upsilon(3S)$ in VMD vertex is dominant in $\gamma p \to \Lambda Z_{bs}^{(')}$, while $\Upsilon(2S)$ is larger than $\Upsilon(1S)$ in $\gamma p \to n Z_{b}^{(')}$. Thus it is reasonable to neglect other bottomonium in VMD for an order-of-magnitude estimation at present. In Fig.~\ref{fig:Zcsphoto}(b), we compare the total cross sections of $\gamma p \to \Lambda Z_{b}^{(')}$ with
only $\Upsilon(3S)$ in VMD vertex to that of $\gamma p \to \Lambda Z_{cs}^{(')}$. It seems that they are in the same level if taking $\mcb (Z_{cs}^{(')} \to J/\psi K)$ = 1.0. Take care that $\mcb (Z_{b}^{(')} \to \Upsilon(3S) \pi) \sim$ 2\% has been included in these plots. Therefore the actual cross sections of $\gamma p \to \Lambda Z_{cs}^{(')}$ would be 1 $\sim$ 2 orders of magnitude smaller than that of $\gamma p \to \Lambda Z_{b}^{(')}$ by considering 1\% $< \mcb (Z_{cs}^{(')} \to J/\psi K)<$ 10\% in various models~\cite{Yang:2020nrt,Wan:2020oxt,Ikeno:2020csu}.

\begin{figure}
  \begin{center}
  {\includegraphics*[width=0.45\textwidth]{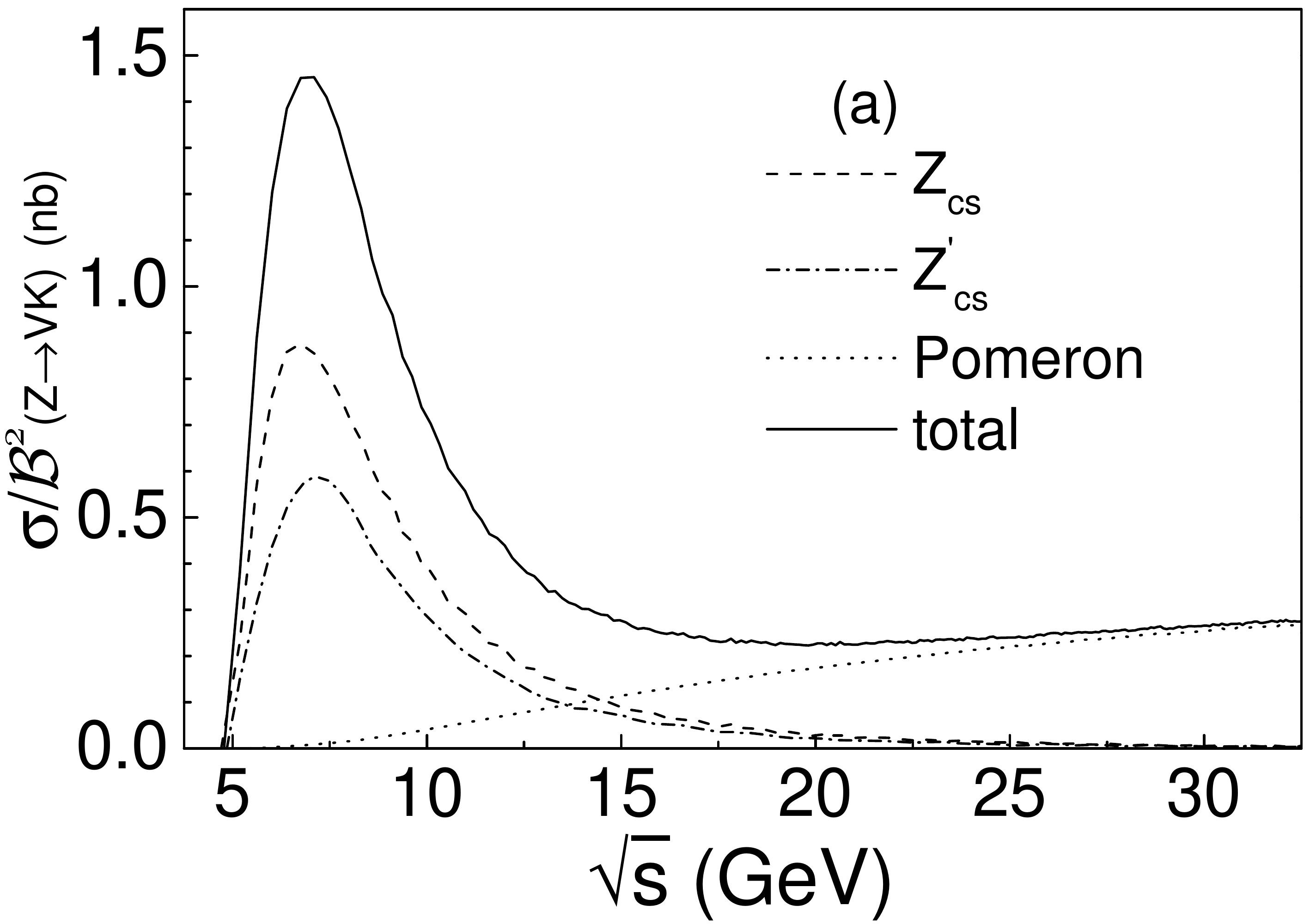}}
  {\includegraphics*[width=0.45\textwidth]{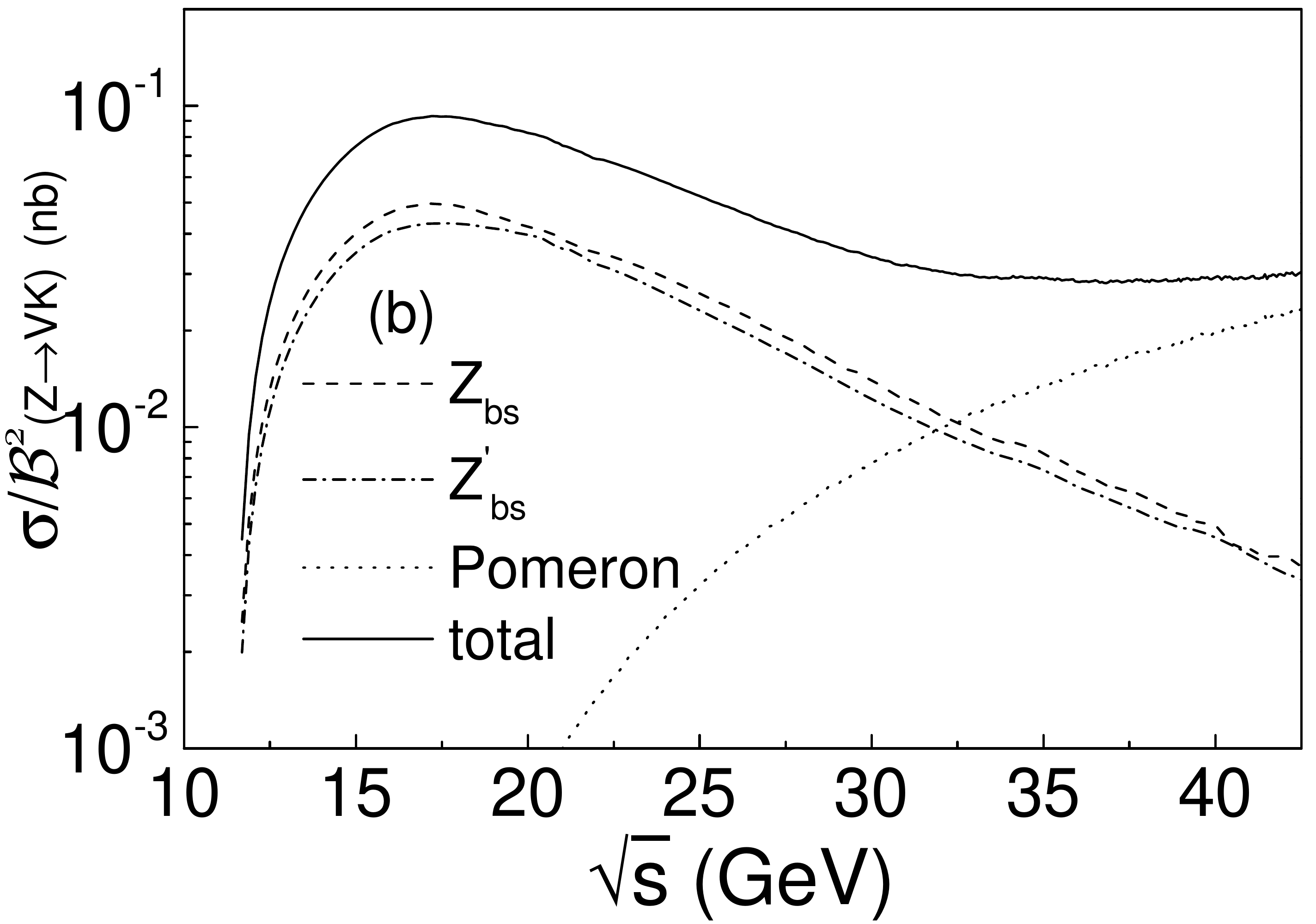}}
    \caption{(a) The total cross section of $\gamma p \to \Lambda J/\psi K$ reaction with the contributions of $Z_{cs}^{(')}$ and Pomeron exchange. (b) The total cross section of $\gamma p \to \Lambda \Upsilon K$ reaction with the contributions of $Z_{bs}^{(')}$ and Pomeron exchange. Note that the vertical axis of (b) is in logarithmic scale.}
    \label{fig:ZcsZbsphoto3}
  \end{center}
\end{figure}

The photoproduction of different $Z_{cs}^{(')}$ states with various $\psi(nS)$ states in the VMD vertex shall be coherently added together when detecting them through $\gamma p \to \Lambda J/\psi K$ reaction. The same is true for $Z_{bs}^{(')}$ photoproduction in $\gamma p \to \Lambda \Upsilon K$. This is not considered by previous study of $Z_{c}$ photoproduction in $\gamma p \to p J/\psi \pi$~\cite{Liu:2008qx,Galata:2011bi,He:2009yda,Lin:2013mka,Wang:2015lwa} mainly because of the unknown relative phases of various amplitudes. To further proceed we have to assume that
these relative phases are zero and various contributions are constructively interfering.
In Fig.~\ref{fig:ZcsZbsphoto3}, we show the cross sections of $\gamma p \to \Lambda J/\psi K$ and $\gamma p \to \Lambda \Upsilon K$ under these assumptions. As stated in Sec.~\ref{sec:Frwk}, the $\sigma(\gamma p \to Z_{cs}^{(')} \Lambda \to J/\psi K \Lambda)$ and $\sigma(\gamma p \to Z_{bs}^{(')} \Lambda \to \Upsilon K \Lambda)$ are excellently approximated by $\sigma(\gamma p \to Z_{cs}^{(')} \Lambda) \mcb(Z_{cs} \to J/\psi K)$ {and $\sigma(\gamma p \to Z_{bs}^{(')} \Lambda) \mcb(Z_{bs} \to \Upsilon K)$}, respectively. The background from Pomeron in $\gamma p \to \Lambda J/\psi K$ in Fig.~\ref{fig:Zbcsphoto3}(b) is small if the cut-off parameters are inhabited from $\gamma p \to p J/\psi \pi$~\cite{Lin:2013mka}. We choose appropriate cut-off in the form factor for Pomeron contribution in $\gamma p \to \Lambda \Upsilon K$, considering that $\sigma (\gamma p \to p \Upsilon)$ is below 0.1 nb when $\sqrt{s}$ is under 100 GeV~\cite{Cao:2019gqo}. It can be concluded that the signal channel is prominent in comparison with the background estimated by Pomeron exchange, however, well below 1 nb and out of reach of the current luminosity design $(2\sim4)\times 10^{33}\text{cm}^{-2}\text{s}^{-1}$ of Electron-Ion Collider in China (EicC)~\cite{CAO:2020EicC}. If the luminosity of EicC increases at least one order of magnitude, these strange states would be hopefully detected in photoproduction reaction by combining with proper construct technique~\cite{Yang:2020eye,Xie:2020niw}. For the proposed Electron-Ion Collider in US (US-EIC) with the luminosity of
$10^{34}\;\text{cm}^{-2}\text{s}^{-1}$ or higher~\cite{Accardi:2012qut}, it would be possible to observe these states.
It shall be further mentioned that the signal of exotic states would be also faded away in the $\gamma p \to \Lambda J/\psi K$ (or $\gamma p \to \Lambda \Upsilon K$) if the destructive interfere of twin states $Z_{cs}^{(')}$ (or $Z_{bs}^{(')}$) is present.

\begin{figure}
  \begin{center}
  {\includegraphics*[width=0.7\textwidth]{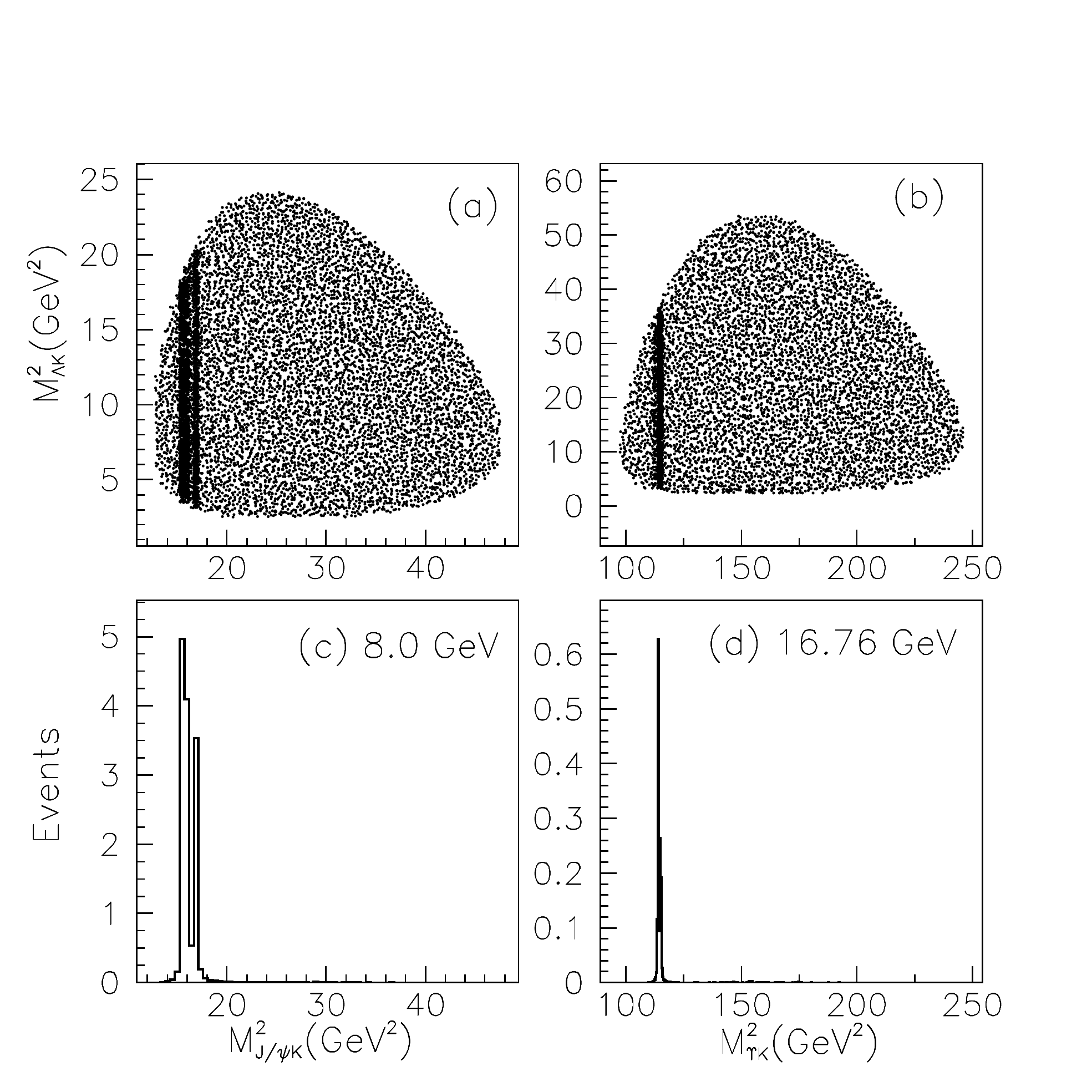}}
    \caption{The Dalitz plots (top) and invariant mass spectra (bottom) of the (a,c) $\gamma p \to \Lambda J/\psi K$ at $\sqrt{s} = $8.0 GeV and (b,d) $\gamma p \to \Lambda \Upsilon K$ at $\sqrt{s} = $ 16.76 GeV.}
    \label{fig:ZcsZbsSpec3}
  \end{center}
\end{figure}

In Fig.~\ref{fig:ZcsZbsSpec3}(a,c), we present the Dalitz plot and $J/\psi K$ invariant mass spectrum of the $\gamma p \to \Lambda J/\psi K$ reaction at $\sqrt{s} = $8.0 GeV. It is hopeful to separate the $Z_{cs}$ and  $Z_{cs}^{'}$ in this channel if enough events are accumulated. In Fig.~\ref{fig:ZcsZbsSpec3}(b,d), we show the Dalitz plot and $\Upsilon K$ invariant mass spectrum of $\gamma p \to \Lambda \Upsilon K$ reaction at 16.76 GeV, which is the optimal energy of the proposed EicC~\cite{CAO:2020EicC}. It can be seen that it is very challenging to distinguish the very narrow $Z_{bs}$ and  $Z_{bs}^{'}$ because the invariant mass $M^2_{\Upsilon K}$ covers wide kinematic range, so a very fine resolution is required.

%

\section{Summary and Conclusion} \label{sec:summ}

Stimulated by the newly observed charged strange hidden-charm state $Z_{cs}$(3985) and the charm-strange states $X_0/X_1$(2900), a full axial-vector heavy quark $Z$ spectroscopy is believed to be emerging.
In this work we investigate the photoproduction of $Z_{cs}^{(')}$ and $Z_{bs}^{(')}$ under the mechanism of VMD and $t$-channel meson exchange since the HQSS has spoken that exotica come with pairs. The $Z_{b}^{(')}$ photoproduction is also presented for comparison.
The signal of resonances would be obvious or obscure depending on constructive or destructive interference of the twin states. If they are constructively enhanced, the cross sections of strange state are around 1 $\sim$ 2 orders of magnitude smaller than that of corresponding non-strange state, as shown by our calculation. Numerically, photoproduction cross sections of $Z_{cs}^{(')}$ and $Z_{bs}^{(')}$ are below 1 nb and 0.1 nb, respectively.
The electroproduction cross sections of these states would be further reduced by two orders of magnitude due to the electromagnetic coupling.
The Dalitz plots and invariant mass spectra indicate that a fine resolution of $J/\psi K$ in $\gamma p \to \Lambda J/\psi K$ and $\Upsilon K$ in $\gamma p \to \Lambda \Upsilon K$ would be prerequisite to identify these narrow states experimentally.
As a result, it is very challenging to search for them at EicC and but possible at US-EIC due to
a higher luminosity. 

Our estimations give a natural order hierarchy for the photoproduction of strange and non-strange states in charm and bottom sector, as already found in the light quark sector, e.g. $pp$ collisions~\cite{Cao:2007md,Cao:2010km} and $\gamma p$ reactions~\cite{Cao:2013psa}. Then we can expect that the cross section of $X_0/X_1$(2900) photoproduction is very tiny, e.g. through $\gamma p \to X_{0/1} \Lambda_c^+ \bar{K}^0$.

Before taking seriously these results, it is worth pointing out that several sources of model uncertainty shall be kept in mind.
First, the mentioned destructive interference of nearby twin states would reduce the cross sections. Second, the VMD model is not careful inspected in the heavy quark sector, as noticed by the previous studies~\cite{Wu:2019adv,Cao:2019kst,Cao:2019gqo}. Third, the cut-off in form factor is not well scaled because of the unavailable data. Thus the search for heavy exotic states through photoproduction would help to examine these aspects on one hand, and gain more insights on the nature of exotic states on the other hand.


\begin{acknowledgments}

We would like to thank F.-K. Guo, X.-H. Liu and M. P. Valderrama for useful discussions. This work was supported by the National Natural Science Foundation of China (Grants Nos. 12075289 and U2032109) and the Strategic Priority Research Program of Chinese Academy of Sciences (Grant NO. XDB34030300).

\end{acknowledgments}

\end{document}